\documentclass[12pt]{article}
\usepackage{hyperref}
\usepackage{graphicx}
\usepackage{rotating}

\usepackage{amssymb}

\def\g{\gamma}

\def\beq{\begin{equation}}
\def\eeq{\end{equation}}
\def\beqn{\begin{eqnarray}}
\def\eeqn{\end{eqnarray}}
\def\ba{\begin{eqnarray}}
\def\ea{\end{eqnarray}}

\def\slash#1{#1\hskip-6pt/\hskip6pt}

\def\l{\langle}

\def\xprim2bar{\overline{x}^{\prime\prime}}

\def\beq{\begin{equation}}
\def\eeq{\end{equation}}

\newcommand{\beqa}{\begin{eqnarray}}
\newcommand{\eeqa}{\end{eqnarray}}

      \let\g=\gamma   
         
      \let\l=\lambda

      \let\g=\gamma   
         
      \let\l=\lambda

\newcommand{\be}{\begin{equation}}
\newcommand{\ee}{\end{equation}}
\newcommand{\bea}{\begin{eqnarray}}
\newcommand{\eea}{\end{eqnarray}}

\usepackage{graphics}
\begin{document}
\setcounter{page}{1}
\vspace{1.0cm}
\begin{center}{\large \bf Windows over a New Low Energy Axion}
\vspace{.17in}

\vspace{.5cm}

{\bf\large Claudio Corian\`{o}$^{1}$$\;$ and  Nikos Irges$^{2}$}

\vspace{.12in}
\vspace{1cm}

{\it  $^1$Dipartimento di Fisica, Universit\`{a} del Salento, and \\
INFN Sezione di Lecce,  Via Arnesano 73100 Lecce, Italy}\\
~\\
{\it
$^2$Department of Physics and Institute of Plasma Physics, \\
University of Crete, GR-710 03 Heraklion, Greece\\}
\end{center}
\vspace{.5cm}

\begin{abstract}

We outline some general features of possible extensions of the Standard Model
that include anomalous $U(1)$ gauge symmetries, a certain number of axions 
and their mixings with the CP-odd Higgs sector. As previously shown, after the mixing 
one of the axions becomes a physical pseudoscalar (the axi-Higgs) that can take the role of 
a modified QCD axion. 
It can be driven to be very light by the same non-perturbative effects that are held responsible for the solution of the strong 
CP-problem. At the same time the axi-Higgs has a sizeable gauge interaction, which is not allowed to the 
Peccei-Quinn axion, possibly explaining the PVLAS results.
We point out that the Wess-Zumino term, typical of these models, can be both interpreted as an anomaly 
inflow from higher dimensional theories (second window) 
but also as a result of partial decoupling of an 
extra Higgs sector (and of a fermion) that leaves behind an effective anomalous abelian theory 
(first window) in a broken St\"{u}ckelberg phase. The possibility that the 
axi-Higgs can be heavy, of the order of the 
Higgs mass or larger, however, can't be excluded.  The potentialities for the discovery 
of this particle and of anomaly effects in the neutral current sector at the LHC are briefly discussed 
in the context of a superstring inspired model (second window), but with results that remain valid also if any of the two possibilities 
is realized in Nature.

\newpage
\end{abstract}
\smallskip

\bigskip

\section{\bf Introduction.}

Extensions of the Standard Model (SM) incorporating a gauge structure with additional abelian $U(1)$'s may 
be of particular relevance in the search for new neutral currents at the LHC and may have profound 
implications in an attempt to uncover the unique extension of the Standard Model that Nature selects.  
Abelian extensions are quite ubiquitous both in Grand Unified Theories and in effective string theories, 
and in general these interactions can be viewed as the low energy remnants of larger symmetries. 
As such, they deserve special attention.
Extra abelian gauge symmetries imply, as we have just mentioned, new neutral interactions 
($Z'$ gauge bosons) and corrections 
to the interactions of the SM electroweak gauge bosons that 
should be quite small, given the accuracy of the electroweak precision data from LEP. 

In this work we intend to underline some simple consequences of the effective low energy theory 
that one obtains under certain assumptions which are, however, 
quite generic if some abelian interactions are present both at the electroweak 
scale and above it; this second scale, that we call $M$, 
can be quite remote from the latter. The appearance of these 
interactions and low energy structures, because of their generality, could be produced by 
different mechanisms, but they have a common low energy imprint: the presence of a pseudoscalar that 
undergoes mixing with the electroweak Higgs so to produce a low energy theory that is unitary, though 
non-renormalizable. This pseudoscalar, that in a first work has been called ``the axi-Higgs'' 
\cite{cik} plays a subtle role in guaranteeing the unitarity of the low energy extension and is characterized by a mass and interactions that are sensitive both to the QCD $\theta$-vacuum but also to additional CP-odd phases of possible scalar 
potentials generated at higher energy and that survive in the low-energy 
theory. In this work we briefly outline the arguments and highlight some of the results, leaving some of the details to a future work. Phenomenological discussions of alternative 
approaches in which the axion-gauge field interaction is modified respect to traditional axion 
models can be found in \cite{Ringwald} \cite{Mirizzi}. 

\section{The first window with a simple model } 

We start from a simple example. A working model that shares the main features discussed above such as a 
St\"{u}ckelberg mass term, an anomalous fermion spectrum and axion-gauge field interactions 
is described by the classical lagrangean 
\beqa
\mathcal{L}_0 &=& |(\partial_{\mu} + i B_{\mu} ) \phi | ^{2} -\frac{1}{4} F_{A}^{2}
-\frac{1}{4} F_{B}^{2}   + \frac{1}{2}( \partial_{\mu} b + 
M_1\ B_{\mu})^{2} -\lambda( |\phi|^{2} - \frac{v^{2}}{2})^{2}   \nonumber\\
&& + \overline{\psi} i \gamma^{\mu} [ \partial_{\mu} +i e A_{\mu}
+ i \gamma^{5} B_{\mu}  ] \psi +\lambda_1 \overline{\psi}_L \phi \psi_R + \makebox{h.c.}
\label{lagrangeBC}
\eeqa
that we call the $A$-$B$ model, with $A$ being vector-like and massless, while $B$ is made 
massive by a combination of the  St\"{u}ckelberg and the Higgs mechanisms. 
The tree-level mass $M_1$, also called ``the St\"{u}ckelberg mass'' combines with the vev of the 
Higgs field to generate the total mass of the $B$ gauge boson which is $\sqrt{M_1^2 + (g_B v)^2}$.
The anomaly of the fermion  is due to its (purely) axial coupling to $B$.
The mixing of the  field $b$ and the Higgs field takes place after 
spontaneous symmetry breaking and is triggered by the Higgs field. 
Probably the simplest way to look at this lagrangean 
is to consider it to be the low energy effective theory of a first breaking, 
driven by a much heavier Higgs ($\Theta$, charged under $U(1)_B$) at a higher scale, whose magnitude appears frozen 
at lower energy. 

Notice that the presence of the St\" {u}ckelberg scalar, that can be thought of as the (surviving) 
phase of $\Theta$, indicates that, before symmetry breaking of the light Higgs, the theory is already in a broken phase (the St\"{u}ckelberg phase). This picture clearly does not give any special 
role to the St\"{u}ckelberg scalar, other than being a CP-odd component of another Higgs field. In this specific interpretation of the Higgs-St\"{u}ckelberg 
system the scale $M$ is directly related to the vev of the heavy Higgs, denoted by $V$. When also the light Higgs takes a vev ($v$), 
we will call the corresponding symmetry phase the Higgs broken phase. 

Both symmetry phases, in principle, could be anomalous, even if the 
original theory was assumed to be anomaly-free. In fact, together with 
the radial component of a heavier Higgs also 
some of the fermions may have been integrated out. The  phase
is then a simple example of an incomplete decoupling of some mother 
theory with a larger symmetry. Clearly both the low energy theory and 
the mother theory need to share a gauged $U(1)_B$ symmetry. 

The left-over fermion, here denoted by $\psi$, makes up the light spectrum.
In the absence of the Higgs field, the axion $b$ has a simple derivative coupling to the $B$ gauge boson, 
and it is a Goldstone mode. In the presence of the vev of a Higgs field, Higgs-axion mixing occurs and one linear combination becomes massless but is now physical. We call this a massless axi-Higgs. Mass corrections to the 
axi-Higgs can be generated by the introduction of phases in a given potential, whose origin, at this point, is left unspecified. We will comment upon this in the following. It is given by 
\beqn
V_{\slash P \slash Q} = \l_2 \left( \phi \, e^{- i q^{}_{B} g^{}_{B} \frac{b}{M^{}_{1}}}   \right) 
+ \lambda_3 \left( \phi \, e^{- i q^{}_{B} g^{}_{B} \frac{b}{M^{}_1} }   \right)^{2} 
+2 \lambda_{4} \left(  \phi^{*} \phi  \right)   \left( \phi  \, e^{- i q^{}_B g^{}_{B} \frac{b}{M^{}_1} } \right)    + \cdots \mbox{c.c.}
\eeqn
with $\l_2$, $\lambda_3$ and $\lambda_4$ suitable parameters. The dots refer to other allowed 
terms, an example of which can be found in \cite{cik}. 
If we denote as $V _{PQ}$ the ordinary Higgs potential (which is phase-independent), 
$V_{\slash P \slash Q}$ introduces a periodic dependence  similar to the breaking of the 
PQ symmetry by non-perturbative effects in the $\theta$-vacuum. 
Then the complete potential considered here can be equivalently written as 
\beqn
V(H, b) = V_{PQ} +  V_{\slash P \slash Q} + V^{*}_{\slash P \slash Q}
\eeqn 
and we require that its minima are located at 
\beq
\langle b \rangle=0 \qquad \langle \phi \rangle =v,
\eeq
which can be achieved by a suitable choice of its free parameters. The CP-odd phases of the scalar 
sector can be rotated into the physical axi-Higgs $(\chi)$ and a Nambu-Goldstone boson $G$, giving
\beqn
\phi_2  &=& \frac{1}{M_B}  (- M^{}_1 \, \chi + q^{}_B g^{}_{B} \, v \, G   )\nonumber   \\
b &=& \frac{1}{M_B} (  q^{}_B g^{}_{B} \, v \,\chi  +  M^{}_1 \, G   ).   
\label{mix}
\eeqn
The phase dependence of the new potential $V_{\slash P \slash Q}$ plays a key role in determining the actual mass of the physical axion. While the expressions of the two linear 
combinations (\ref{mix}) remain true also for a vanishing $V_{\slash P \slash Q}$, 
when it is present, there are some important mass corrections generated as well. 

More precisely, for a non-vanishing $V_{\slash P \slash Q}$ the scalar mass matrix 
has one zero eigenvalue corresponding to the Goldstone boson $G$ and a non-zero 
eigenvalue corresponding to a physical axion field $- \chi -$ with mass
\beqn
m^2_{\chi} =  - \frac{1}{2} c_{\chi} v^2 \left[  1 + \frac{q^2_B g^{2}_{B} v^2}{M^2_1} \right] = - \frac{1}{2} c_{\chi} \, v^2 \, 
\frac{M^{2}_{B}}{M^{2}_{1}}. 
\eeqn 
The mass of this state is positive if $c_{\chi} < 0$, with
\beqn
c_{\chi} = 4 \left( \frac{\l_2}{v^3} + \frac{4 \lambda_3}{v^2} + \frac{2 \lambda_4}{v} \right).
\eeqn
The massless axi-Higgs is obtained by sending the parameters of $V_{\slash P \slash Q}$ to zero, 
that is $c_\chi\to 0$.
Notice that this parameter plays an important role in establishing the size of the mass of $\chi$, 
and encloses also the corrections to the standard Higgs potential induced by the phase 
dependent potential. Notice also that at this point we obtain different 
scenarios depending on the size of these parameters. 
In particular they could be generated non-perturbatively similarly 
to the case of the PQ axion, due to the presence of 
the instantons in the QCD $\theta $-vacuum \cite{Wilczek} 
and in that case they would be naturally small \cite{Sikivie}. 
A scenario that would guarantee a large vev $V$ for the heavy Higgs $\Theta$ 
while not allowing a phase to develop in the scalar potential, is one where the
effective scalar potential originates from a supersymmetric D-term in the presence
of a non-zero Fayet-Iliopoulos term. Notice that this possibility leaves open a (third) window
towards a connection with the Froggatt-Nielsen mechanism \cite{Proceed}.

\subsection{Gauge interactions of the axi-Higgs}

The theory described above suffers from gauge anomalies, true or apparent, 
and there could be, at this stage, questions about its implementation at the level of effective field 
theories. Clearly, it is reasonable to expect that the main shortcoming of an 
effective anomalous theory is its non-renormalizability, 
while its unitarity should be preserved. 

In the absence of a low-energy Higgs-axion mixing 
from a scalar potential (i.e. $V_{\slash{P} \slash{Q}}\to 0$) and in the absence of Yukawa couplings 
(vanishing fermion masses) we can show unitarity of the model by a direct analysis with ease. 
In the presence of Yukawa couplings in a broken Higgs phase the same computation is slightly 
more involved but also goes through.
 
The tests can be performed by analyzing the cancelation of the gauge dependence of the 
gauge fixed action in a specific (non-unitary) gauge.

This can be more easily shown in 
the $R_\xi$ gauge, using the BRST symmetry of the full effective action and the validity of the corresponding Slavnov-Taylor identities. However, prior to gauge fixing and in the presence of 
a gauge field-axion mixing, the anomalous effective action, 
defined (only) as the classical action plus the anomalous triangle diagrams - in our case these are the 
($AAB$) and the ($BBB$) vertices -  can be rendered gauge invariant by suitable counterterms.  

A wide body of literature 
on anomalies, in the past, has tried to answer this puzzling question. It has been proposed, 
just from a field theoretical ground, that an anomalous theory can be ameliorated by a 
suitable Wess-Zumino term, that introduces an axion. These arguments have been brought up (quite long ago) 
in theories in which a non-dynamical $\theta$-term \cite{GrossJackiw,Faddeev} could improve the 
analysis of the effective theory.

The basic conclusion
of these investigations was that a Wess-Zumino term does not ``cancel'' the anomaly (see the discussion in 
\cite{Preskill}), but ameliorates 
the behavior of the theory allowing a perturbative power counting. Also, the addition of non-local 
counterterms that do restore gauge invariance \cite{Krashnikov} are 
not compatible with an effective renormalizable theory. We are going to illustrate how 
this power counting goes. 

\subsection{Restoring gauge independence} 

\begin{figure}[t]
{\centering \resizebox*{15cm}{!}{\rotatebox{0}
{\includegraphics{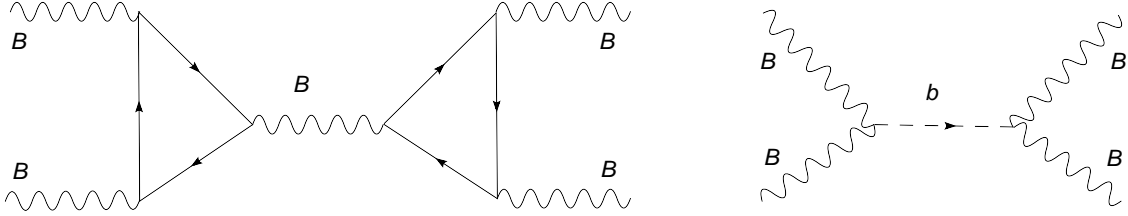}}}\par}
\caption{ Diagrams relevant for the analysis of the 
gauge independence in the $R_\xi$ gauge for a B exchange.}
\label{unitaritycheck}
\end{figure}

The introduction (just on a field theoretical basis) of a WZ term, a dimension-5 operator 
that spoils renormalizability, is the price to pay in order to be able to discuss 
the gauge-independence of the S-matrix amplitudes of the model. This is illustrated in 
Fig.~\ref{unitaritycheck}, where the diagrams are analyzed in the unitary gauge. 
Both diagrams are needed to remove the gauge dependence of the first diagram, the second one being purely 
gauge dependent. Had we not 
introduced this term, we would be forced to cancel completely the gauge dependence of the first diagram just by going to the unitary gauge. In the unitary gauge $b$ is set to vanish and the second diagram disappears. In the presence of a Higgs, however, the same diagram does not disappear, since the unitary gauge choice does not eliminate the exchange of a physical axion, which corresponds to the exchange of a massless or massive pole depending 
on the absence/presence of a phase dependent potential. In this case the axion is rotated 
as in Fig.~\ref{gs}. 

\subsection{Unitarity}

\begin{figure}[t]
{\centering \resizebox*{15cm}{!}{\rotatebox{0}
{\includegraphics{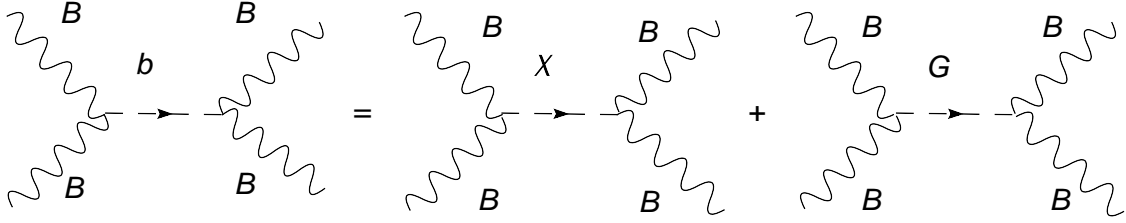}}}\par}
\caption{Projection of the axion b over the axi-Higgs and the Goldstone in the presence of Higgs-axion mixing.}
\label{gs}
\end{figure}

\begin{figure}[t]
{\centering \resizebox*{13cm}{!}{\rotatebox{0}
{\includegraphics{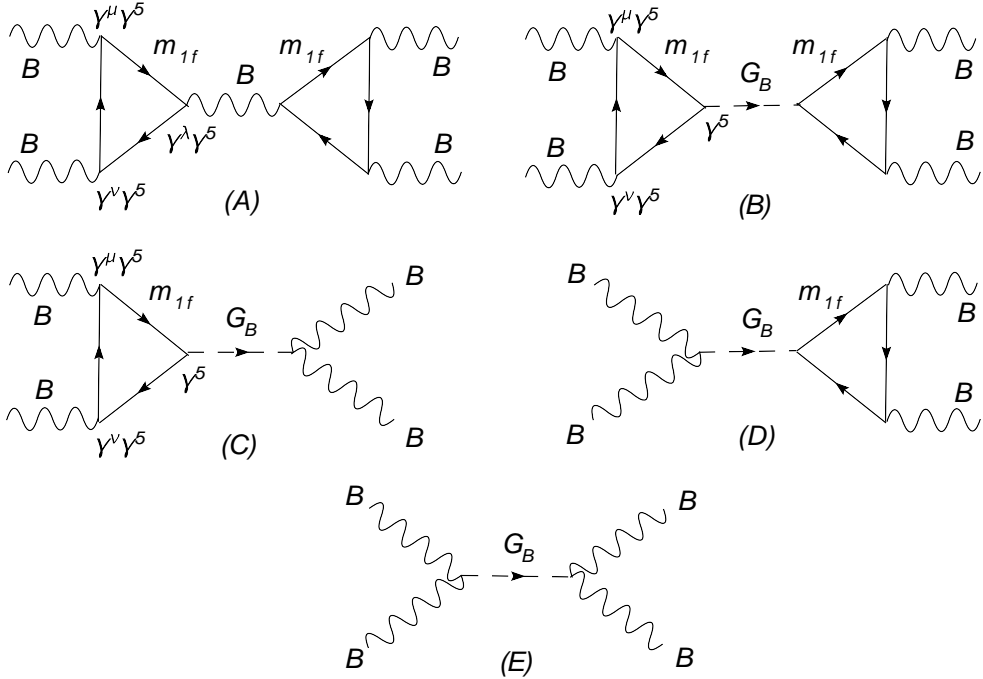}}}\par}
\caption{Gauge dependence cancelation after spontaneous symmetry breaking.}
\label{brokenphase}
\end{figure}
%

We are now going to show that the theory constructed out of the St\"{u}ckelberg term and with/without a Yukawa 
coupling is unitary in the broken Higgs phase - the presence of the WZ term being necessary - 
by a direct computation.  

The simplest approach to test a non-unitary behavior of a theory is to look for the presence of spurious 
s-channel poles in a variety of scattering amplitudes involving the anomalous diagrams of the 
model as in Bouchiat, Iliopoulos and Meyer \cite{Iliopoulos}.
The proof is quite simple and can be identified by following the patterns of cancelations of the gauge dependence in s-channel exchanges such as those in Fig.~\ref{unitaritycheck}. 

Moving to the same analysis in the broken Higgs phase, a quick look at Fig.~\ref{brokenphase} 
may help in identifying the same pattern in the 
presence of Yukawa couplings. The exchange of the axion $b$ 
is rotated into the Goldstone and into the 
(gauge independent) physical axi-Higgs (not included in Fig.~\ref{gs}). The pattern of cancelations 
then proceeds very similarly to the case of the SM and can be shown explicitly.  We have 
not used any additional counter-term of Chern-Simons (CS) type to achieve this result, though, in principle, the theory may allow for terms of this type 
\cite{cik,Bianchi}. In this case a Ward identity on the $A$ gauge boson 
determines the parametrization of the anomalous triangle diagram (which is defined modulo a momentum 
shift) so that the $A$-vector current is conserved. This condition, in fact, is a defining Ward identity on the model, corresponding to gauge invariance under $A$-gauge transformations. 
Similar Ward identities need to be satisfied if certain symmetries (residual gauge symmetries) 
are left over (such as $U(1)_{\makebox{em}}$ gauge invariance) after symmetry breaking. 

\subsection{The role of the Chern-Simons interactions} 

We have seen, by a specific example, that the WZ term is more than just a gauge artifact in the low energy 
effective action and plays a considerable role in restoring the unitarity of the theory even 
in a physical gauge, since the axi-Higgs contributes to a given S-matrix element in a gauge 
independent way. Besides the WZ interaction, the model, in principle, allows direct gauge interactions 
mediated by Chern-Simons terms. While in more general models involving extensions of the SM that include more than a single anomalous $U(1)$ these interactions are part of the effective action, in the case of the model 
that we have illustrated this role is lost. It is however interesting to re-analyze the same 
pattern of cancelations in a more general case. 

Therefore, we consider the following modification on the $AB$ model where the CS interactions are generically introduced as possible counterterms, together with WZ terms, in the 1-loop effective action, which is given by 
\beqn
{\mathcal L} = {\mathcal L_0} + {\mathcal L_{WZ}} + {\mathcal L_{CS}},
\eeqn
where ${\mathcal L_0}$ is the starting lagrangean, but in particular we focus on 
\beqn
{\mathcal L_{WZ}} =  \frac{C^{}_{AA}}{M} b F^{}_{A} \wedge F^{}_{A} +\frac{C^{}_{BB}}{M} b F^{}_{B} \wedge F^{}_{B}, 
\eeqn
which corresponds to the WZ term (or Green-Schwarz in the string language)
and on the term
\beqn
{\mathcal L_{CS}} =  d^{}_{1}B^{\mu} A^{\nu} F^{\rho \sigma}_{A} \epsilon_{\mu \nu \rho \sigma} = d^{}_{1}BAF^{A},
\eeqn
which denotes the gauge variant CS interaction,
where gauge invariance fixes the unknown coefficients via the specific relations 
\beqa
C^{}_{AA} &=& \left( - \frac{d^{}_{1}}{2} + \frac{i}{2!}a^{}_{3}(\beta)\frac{1}{4}  \right)\frac{M}{M^{}_{1}} \nonumber \\
C^{}_{BB}&=& \frac{i}{3!} \frac{a_n}{4}\frac{M}{M_1}.
\label{GSexplicit}
\eeqa
One important comment is in order. It is 
of considerable importance to observe that in these models the size of the 
interaction of the axi-Higgs to the gauge fields, $1/M$, 
remains unrelated to the mechanism that generates its mass. 
This is an important variant compared to the standard PQ invisible axion, where they are both controlled by the same parameter $1/f_a$ and both the mass term and the axion gauge interactions are equally suppressed 
by the same scale ($f_a\sim 10^{10}\, \makebox{GeV}$). It is by now well known that this point determines a disagreement 
between the PVLAS result \cite{PVLAS} and its interpretation as a traditional axion \cite{Masso}.

In the equations above, $\beta$ is a shift parameter relevant for the distribution of the anomalies in the $AAB$ diagram, 
which in this case are given by 
\beqa
a_1(\beta)&=& a_2(\beta)=-\frac{i}{8\pi^2} - \frac{i}{4 \pi^2}\beta \nonumber \\ 
a_3(\beta) &=& -\frac{i}{4\pi^2} + \frac{i}{2 \pi^2}\beta, \nonumber \\ 
\label{a12}
\eeqa
while the $BBB$ triangle is shift-independent as a result of Bose symmetry. In the equation above the 
distribution of the partial anomalies is such that $a_1=a_2$ and $a_1 + a_2 + a_3=a_n$ 
with $a_n=i/(2 \pi)^2$ being the total anomaly. 

\begin{figure}[t]
{\centering \resizebox*{14cm}{!}{\rotatebox{0}
{\includegraphics{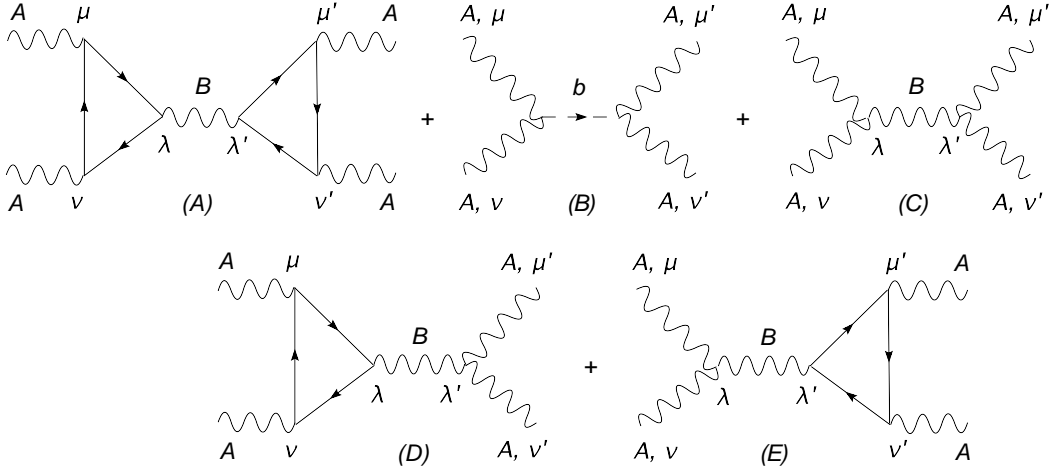}}}\par}
\caption{Diagrams involved in the unitarity analysis with CS interactions.}
\label{chern}
\end{figure}

It can be shown that $\beta$ in (\ref{a12}) can be fixed by the Ward identity on $A$, 
giving a conserved vector current (CVC), or can be left arbitrary at the price of introducing a CS interaction. Both approaches give, though, the same physical result. In our parametrization the CVC 
condition that fixes the two divergent amplitudes of the anomaly diagram 
is obtained for $\beta=-1/2$. As we have already mentioned, in general, external Ward identities 
could bring these two invariant amplitudes to a paramaterization which differs 
from Rosenberg's original one \cite{Rosenberg} and that are 
typical of a given theory of this type.\footnote{Notice that different 
parameterizations, in this case, are not just connected via Schouten's relation.}. However, 
as far as (\ref{GSexplicit}) is observed, a diagrammatic cancelation holds, as one can show by 
a direct computation and as illustrated in Fig.~\ref{chern}. CS interactions, present in Fig.~\ref{chern}, in this case, can be ``absorbed'' 
into a standard definition of the (BAA) triangle diagram with a CVC condition on A.

%
\begin{figure}[t]
{\centering \resizebox*{14cm}{!}{\rotatebox{0}
{\includegraphics{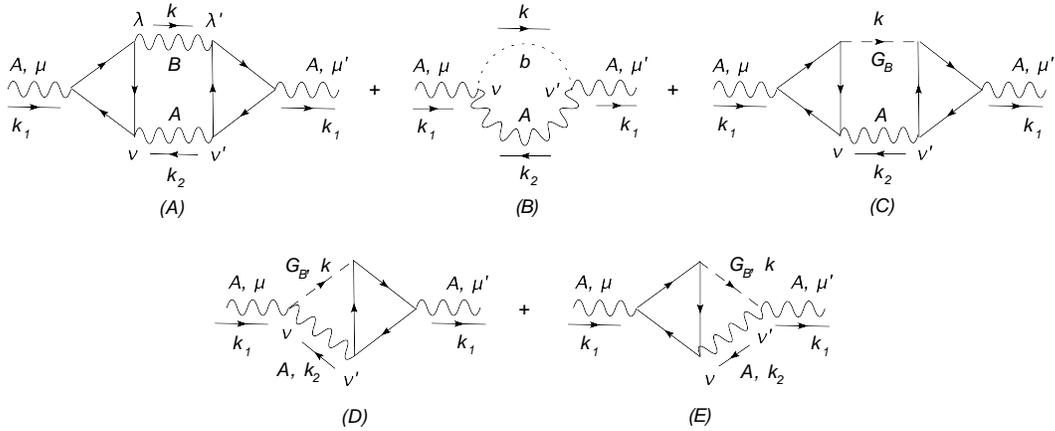}}}\par}
\caption{Diagrammatics for the cancelation of the gauge dependence in the self-energy of $A$}
\label{self}
\end{figure}

\section{ The second window: effective models from string theory (MLSOM and such)} 
As we have mentioned, if one of the possible pictures that points toward a modified axion is 
that of partial decoupling, a second view is related to the higher dimensional origin of the 
anomaly. This second picture, probably less economical compared to the first, can't be ruled out 
either and a growing literature \cite{Kiritsis} on the subject of intersecting brane models 
testifies of this broad interests. Similar pictures, motivated by the study of various models of anomaly cancelations involving brane scenarios with extra dimensions and an anomaly 
inflow, have also been formulated \cite{Hill, Quiros}. These models are characterized 
by extended anomalous abelian gauge structures.

Abelian extensions of the SM can be motivated exactly as we have discussed in the introduction. We could easily entertain the idea that a massive axion, whose mass is unspecified, appears in the spectrum. In the 
$\theta$-vacuum the mass of this axion can be very small and be characterized by an interaction to the fermion sector that is proportional to the fermion mass. If we neglect every 
additional phase dependent potential - as we lower the energy scale at which 
we resolve the theory - except for the instanton corrections from the $\theta$-vacuum, 
the low energy axion develops a mass which is of the same order as the standard Peccei-Quinn axion 
\cite{PQ}.
One important property to check is the exact masslessness of the left-over symmetry 
(in our simple example this is the masslessness of A) and the gauge independence of the corresponding 
corrections, which follow the patterns of cancelation described diagrammatically 
in Fig.~\ref{self}. The disappearance of the spurious s-channel poles is obtained by combining the 
diagrams in a specific pattern dictated by a loop expansion. For instance, the vertex counterterms in this expansion counts one power of $\not{h}$, and, as one can check, their sum is unambiguously defined 
and gauge independent.

We consider the case of an 
$SU(3)_C\times SU(2)_W\times U(1)_Y\times U(1)_B$ model, with an anomaly-free hypercharge 
(traceless with the ordinary generators of the SM), the SM fermion content and one shifting axion, $b$. We denote with $Y_B$ the extra generator of the B gauge boson.
The only anomalous contributions are coming from the $Y$-$Y_B$ mixing which are canceled by external 
Ward identities. Also in this case direct CS interactions are absent.  
In this case, as in the previous case, we have already shown that the axion must be rotated 
in the broken phase and expressed in terms of a physical axi-Higgs and two Nambu-Goldstone modes 
\bea
b &=& c\, \chi +  c_1 G_1^0  +  c_2 G_2^0,     \nonumber\\
\label{rot12}
\eea
where $c$ and $c_i$ are dimensionless,
computable but model dependent coefficients. At a second stage, the two goldstones can be expressed in terms 
of $G_Z$ and $G_{Z'}$, the corresponding goldstones of the two neutral massive gauge bosons.

%
\begin{figure}[t]
{\centering \resizebox*{14cm}{!}{\rotatebox{0}
{\includegraphics{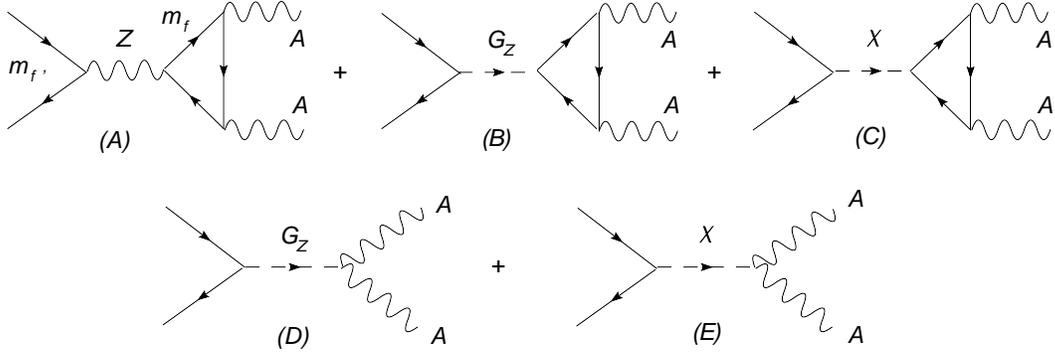}}}\par}
\caption{Contributions to two-photon production  from $q \bar{q}$ annihilation}
\label{ampiezza3}
\end{figure}

%

\begin{figure}[t]
{\centering \resizebox*{15cm}{!}{\rotatebox{0}
{\includegraphics{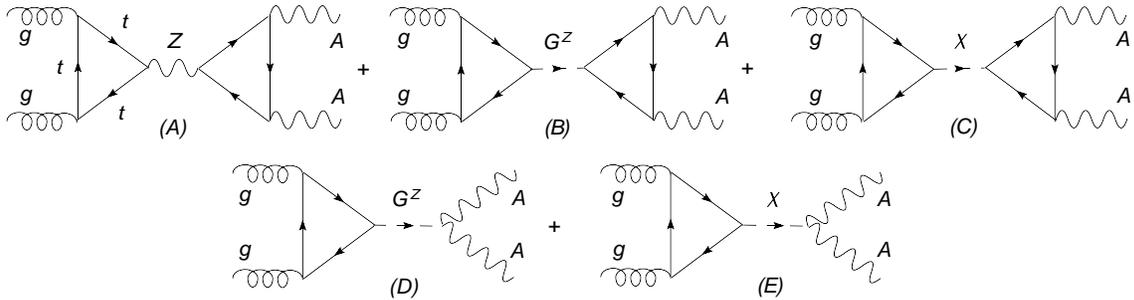}}}\par}
\caption{$Z$ exchange from anomalous gluon fusion}
\label{gluoncorretto}
\end{figure}
\begin{figure}[t]
{\centering \resizebox*{12cm}{!}{\rotatebox{0}
{\includegraphics{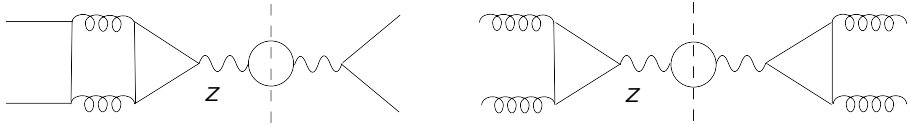}}}\par}
\caption{O($\alpha_s^2$) (NNLO) corrections in an anomalous Drell-Yan cross section  }
\label{drellyan}
\end{figure}

\subsection{The axion and anomaly effects at the LHC}

Moving our analysis toward possible discoveries of anomaly effects at the LHC, here we briefly discuss 
the role played by the axi-Higgs and its mixing in simple partonic processes. We have selected one of the simplest 
processes that can be studied in this new framework, which is the $Z\gamma \gamma$ vertex.
Also in this case, a detailed unitarity analysis of the model can be pursued but details will be given 
elsewhere. We show in Figs. \ref{ampiezza3} and \ref{gluoncorretto} the set of contributing diagrams that are 
responsible for an anomalous decay of the $Z$ gauge boson produced from a quark-antiquark collision. 
The decay channel shown here is the 2-photon decay. Similar processes 
involving gluons could be taken into consideration.
The process occurs with a virtual $Z$ and two real photons, while 
for an on-shell $Z$ it can take place with one direct and one resolved photon in the final state. 
A similar decay channel may involve gluons or be mediated by the only $Z'$ 
gauge boson of this extension. This is generated by the mixing of $Y,Y_B$ with the neutral component of $SU(2)$. The role played by the triangle anomaly in these type of processes is quite 
subtle, and is related both to massless and to massive fermion effects, 
as opposed to the SM, where the anomalous contributions 
are due only to the difference in mass of each generation of fermions.

In order to simplify the treatment, it is convenient 
to introduce some notation. We define the chiral asymmetries for the traces of the generators 
$X_i$, with $i$ any latin index 
labeling both the abelian and the non-abelian cases. All the fermions (f) on which we sum 
are assumed to be in the same generations of the SM and we will neglect 
the sum over different generations just for simplicity. These asymmetries are given by the differences between the left (L) and (R) chiral components for each fermion

\beq
\theta^{(SM) f}_{XYZ}=  
Tr_f(X_i^L X_j^L X_k^L)_{SM} - Tr_f(X_i^R X_j^R X_k^R)_{SM}.
\eeq
We adopt a similar definition for generators with charges in the extended model 
(an example is the MLSOM of \cite{cik})
\beq
\theta^{(MLSOM) f}_{XYZ}=  
Tr_f(X_i^L X_j^L X_k^L)_{MLSOM} - Tr_f(X_i^R X_j^R X_k^R)_{MLSOM},
\eeq
the simplest of them having a single anomalous $U(1)$, (denoted as $U(1)_B$, beside the hypercharge. In this case, the anomalous contributions involve the 
$B\,W\,W$ and the $B\,Y\,Y$ triangles, where $Y$ is the hypercharge, which contribute with non vanishing traces, 
while for the remaining contributions we have the usual conditions  

\beq
\sum_f \theta^{(SM)f}_{X_i X_j X_k}=0,
\eeq
which hold only in the case of the generators of the SM. The derivation of 
the final expression is rather involved, however the crucial step in the derivation is to separate the massive ($\Delta(m_f)$) from the massless 
($\Delta(0)$) contributions of the anomalous diagram ${\Delta}^{\lambda\mu\nu}(m_f\neq 0)$ as

\beq
\Delta^{\lambda\mu\nu}(m_f)=\Delta^{\lambda\mu\nu}(m_f\neq 0)-\Delta^{\lambda\mu\nu}(m_f=0)
\eeq
and by defining $\Delta^{\lambda\mu\nu}(0)\equiv\Delta^{\lambda\mu\nu}(m_f=0)$,
after some manipulations we are left with the following expression of the $Z\g\g$ amplitude in the physical basis in the broken (Higgs-St\"uckelberg) phase
\ba
A^{\lambda\mu\nu}_{Z\gamma \gamma}&=&\underline{\Delta}^{\lambda\mu\nu}(m_f)+\underline{\Delta}^{\lambda\mu\nu}(0) \nonumber \\
&=&  \underline{\Delta}^{\lambda\mu\nu}(m_f)+\frac{1}{4}\sum_f\left[Q_B^{L}Q_Y^{L\,2}-Q_B^{R}Q_Y^{R\,2}\right]\Delta_{AVV}^{\lambda\mu\nu}(0)\,R^{BYY}
\nonumber\\
&&\hspace{2cm}+\frac{1}{4}\sum_f Q_B^{L\,f} (T^{3\,f}_{L})^2 \Delta_{AVV}^{\lambda\mu\nu}(0)\,R^{BWW}, 
\label{ecco}
\ea
where 
\beqa
R^{BYY}&=&(O^{A}_{Y\gamma})^{2}(O^{A})_{B Z}\nonumber\\
R^{BWW}&=&(O^{A}_{W \gamma})^{2}(O^{A})_{B Z}
\eeqa
describe the rotation from the interaction to the mass eigenstates of the 
model \cite{cik}. The massive contributions $\underline{\Delta}(m_f)$ contain 
suitable combinations of all the generators, while $\underline{\Delta}(0)$ is the new massless part of the amplitude which is absent in non-anomalous abelian interactions. This second contribution can be given in terms of standard 
AVV diagrams, as shown in \ref{ecco}, having absorbed some generalized Chern-Simons terms in the amplitude. 
These two amplitudes both satisfy 
the Landau-Yang theorem for the on-shell decay of the $Z$ into two massless photons, which has to vanish. 
The analysis, here shown for a single $U(1)$, can be extended to models 
with more anomalous U(1)'s and shifting axions, following a similar 
approach, along the lines of \cite{cik}. 

\subsection{A brief unitarity analysis}
It is interesting to analize the structure of the various contributions which 
involve these anomalous vertices in the case of neutral currents. 
Once more we reconsider Fig.~\ref{ampiezza3}.
We start from the annihilation of a quark antiquark pair in the parton model.
Notice that, for light quarks, the contributions in Fig.~\ref{ampiezza3} simplify, since the goldstones and the axi-Higgs 
decouple from the fermions (for the sake of clarity, we have included a mass $m_f$ on each diagram 
to emphasize the presence of a mass dependent coupling). In this case 
the only contribution in the quark annihilation channel is due to 
diagram (A). Similar anomalous channels, related to $Z$ production, are shown instead 
in Fig.~\ref{gluoncorretto}.
The production of an axi-Higgs also takes place mediated by fermions in the initial state, similarly to the Higgs. We remark that the coupling of the axi-Higgs to the fermions is similar to the Higgs 
coupling. Corrections to the SM rates of similar processes are due both to the structure of the 
anomaly traces in each {\bf AAA} or {\bf AVV} (Axial or Vector triangle diagram), to the mass 
differences of the fermions in a given generation, and to the presence of direct Wess-Zumino 
interactions, which are however suppressed by powers of $1/M$. For instance, a process characterized by 
direct production via gluon fusion $gg\to Z $ not mediated by a heavy fermion in the initial state would similarly be suppressed by the same factor. Analysis of this type can be carried out in a systematic way.
\subsection{``Anomalous'' Drell Yan at NNLO} 
A final comment concerns the possibility of detecting new effects in neutral current exchanges 
in the Drell-Yan process. The simplest example amounts to an anomalous 
contribution that starts at order $\alpha_s^2$, which is clearly 
neglected in typical SM computations, due to the vanishing anomaly traces. We show 
the two relevant diagrams in Fig.~\ref{drellyan}. They correspond to the interference of the two loop $q \bar{q}$ 
annihilation into a lepton pair with the analogous channel at tree-level, and the 
interference of the two gluon fusion diagrams, which is also of the same order. 
In this case, most of the NNLO QCD analysis performed in the last few years 
will be crucial to achieve sufficient precision on the Z resonance in order to extract 
some essential information on this anomalous process. 

\section{Conclusions} 

We have illustrated some of the physical consequences that are inevitable when an 
anomalous abelian gauge interaction appears at low energy. The presence of 
a new particle, the axi-Higgs, whose mass is the result both of electroweak symmetry breaking and of the phase 
dependent potential(s) generated in the process of breaking of a symmetry 
at higher energy, shows a direct gauge interaction of PQ type with some specific 
features that are clearly missing from the Standard Model but also in the case of the traditional PQ axion model. 
We have argued that the effective interaction of the axi-Higgs to the gauge bosons can be 
generated, apart from a variety of geometrical constructions performed in the context of 
string theory or by anomaly inflow mechanisms from extra dimensions, in a more economical way
by simply integrating out some heavy degrees of freedom (Higgs/fermions) of an anomaly-free 
theory at a higher energy. The model can be realized by two Higgs fields, 
one heavy and one light and an anomaly-free $U(1)$ extra gauge interaction, 
broken by the heavier Higgs at a higher 
scale. The same scale is responsible for the suppression of the interaction of the 
pseudoscalar with the gauge fields at low energy. The mass of the pseudoscalar 
and its gauge interactions are, therefore, unrelated.
This construction renders the Wess-Zumino term simply a low-energy 
manifestation of incomplete decoupling of a heavier Higgs. Analysis of decoupling of a chiral 
fermions are available from the previous literature \cite{Feruglio} and so are the attempts 
to look at variants of the PQ axion \cite{Sanghyeon}, while a heavy axion that mixes with the 
Higgs has also been considered \cite{Rubakov}. On the other hand the St\"{u}ckelberg mechanism has 
 also received a renewed attention \cite{Kors}.
Here, our attempt has been to combine several of these ideas into a fruitful form and we have 
discovered a central avenue: the crucial role played by the WZ term, generated by more economical or by more ambituous constructions. 
We have shown that, in all these cases, the effective theory that is generated 
is non-renormalizable but built so that gauge invariance and unitarity are preserved. 
More details of this analysis will be presented in a forthcoming work.
\vspace{.5cm}

\centerline{\bf Acknowledgements}
We thank Marco Guzzi and Simone Morelli for collaborating to this analysis. 
We thank Elias Kiritsis for collaboration on a previous work that has triggered this analysis. 
We thank P. Anastasopoulos, M. Bianchi, Marco Roncadelli, Theodore Tomaras 
and Cosmas Zachos for discussions. 
The work of C.C. was supported (in part) by the European Union through the Marie Curie Research and Training Network ``Universenet'' (MRTN-CT-2006-035863). 
He thanks the Physics Department at the University of Crete and in particular Theodore Tomaras for the kind hospitality. 
N.I. was partially supported by the Research Program 
PYTHAGORAS II of the Greek Ministry of Education.


\end{document}